\documentclass[aps, showpacs, showkeys,nofootinbib,floatfix]{revtex4}

\usepackage{amssymb}
\usepackage{amsmath}
\usepackage{graphicx}
\usepackage[dvipdfm,colorlinks,citecolor=blue,urlcolor=blue]{hyperref}

\begin{document}

\title{Energy conditions in $f(R,L_{m})$ gravity }

\author{Jun Wang\footnote{wangjun\_3@126.com}, Kai Liao}

\affiliation{Department of Astronomy, Beijing Normal University,
Beijing 100875, China}

\begin{abstract}
In order to constrain $f(R,L_{m})$ gravity from theoretical aspects,
its energy conditions are derived in this paper. These energy conditions given by us are quite general and can be
degenerated to the well-known energy conditions in General Relativity and $f(R)$ theories of gravity with arbitrary coupling, non-minimal coupling and non-coupling between matter and geometry,
respectively, as special cases. To exemplify how to use these energy conditions
to restrict $f(R,L_{m})$ gravity, we consider a
special model in the FRW cosmology and give some corresponding results by using astronomical observations.

\end{abstract}


\maketitle

\section{$\text{Introduction}$}
Recent observations of type Ia supernovae (SNe Ia)\cite{1} indicate that the expansion of the Universe
is accelerating at the present time. Moreover, combination of the observational data of the large scale structure (LSS)\cite{2}, the Cosmic Microwave Background Radiation (CMB)\cite{3} and
the baryon acoustic oscillation (BAO), we know that our Universe is spatially flat
and the cosmic budget of energy is as follows: usual baryon matter occupies
about $4\%$, dark matter occupies about $23\%$ and dark energy
occupies about $73\%$. It is reasonable to believe that the acceleration of the Universe is probably driven by dark energy which is
an exotic component with negative pressure.
Several candidates for dark energy have been extensively
investigated\cite{4}. Unfortunately, up to now a satisfactory
answer to the questions of what dark energy is and where it came from has not yet
been obtained.

Alternative to dark energy, the late-time acceleration of the Universe can be explained by modified theories of gravity\cite{5}.
There are numerous ways to deviate from the Einstein's theory of General Relativity (GR). Among these theories, $f(R)$ theories of gravity (see, for instance, Ref.\cite{6} for
reviews) are very competitive. Here $f(R)$ is an arbitrary function of the Ricci scalar
$R$. One can add any form of $R$ in it, such as $1/R$\cite{7}, $\ln
R$\cite{8}, positive and negative powers of $R$\cite{9},
Gauss-Bonnet invariant\cite{10}, etc. It is worth stressing that
considering some additional conditions, the early-time inflation and
late-time acceleration can be unified by different roles of
gravitational terms relevant at small and at large curvature. Considering the matter-geometry coupling, models of $f(R)$ gravity with non-minimal coupling\cite{11}
and arbitrary coupling\cite{12} have been proposed. Besides them, a more general model of $f(R)$ gravity, in which the Lagrangian density takes the form of $f(R,L_{m})$,
has recently been proposed\cite{13}, and usually called $f(R,L_{m})$ gravity.
Here $f(R,L_{m})$ is an arbitrary function of the Ricci scalar and of the Lagrangian density
corresponding to matter. The astrophysical and cosmological implications of these mentioned models
have been extensively investigated in Ref.\cite{14}.

Since many models of $f(R)$ theories of gravity have been proposed,
it gives rise to the problem of how to constrain them from
theoretical aspects. One possibility is by imposing the so-called
energy conditions\cite{15}. As is well known, these energy
conditions are used in different contexts to derive general results
that hold for a variety of situations\cite{16,17,18,19,20}. Under
these energy conditions, one allows not only to establish gravity
which remains attractive, but also to keep the demands that the
energy density is positive and cannot flow faster than light. In
order to make $f(R,L_{m})$ gravity contain the above mentioned
features, we derive and discuss its energy conditions in this paper.

This paper is organized as follows. The fundamental elements of
$f(R,L_{m})$ gravity are given in section 2. In section 3, the
well-known energy conditions, namely, the strong energy condition
(SEC), the null energy condition (NEC), the weak energy condition
(WEC) and the dominant energy condition (DEC), are derived in
$f(R,L_{m})$ gravity. For getting the first two energy conditions,
the Raychaudhuri equation which is the physical origin of them is
used. It is worth stressing that from the calculation, we find that
the equivalent results can be obtained by taking the transformations
$ \rho \rightarrow \rho^{eff}$ and $p\rightarrow p^{eff}$ into
$\rho+3p\geq0$ and $\rho+p\geq0$, respectively. Thus by extending
this approach to $\rho-p\geq0$ and $\rho\geq0$, the DEC and the WEC
in $f(R,L_{m})$ gravity are obtained. In order to get some insights
on the meaning of these energy conditions, we apply them to a
special model with $f(R,
L_{m})=\Lambda\exp(\dfrac{1}{2\Lambda}R+\dfrac{1}{\Lambda}L_{m})$.
In section 4, by using the parameters of the deceleration $(q)$, the
jerk $(j)$ and the snap $(s)$, the energy conditions of $f(R,L_{m})$
gravity are rewritten. Furthermore, by taking the rewritten WEC and
present astronomical observations and considering $L_{m}=-\rho$ and
$L_{m}=p$\cite{21}, respectively, we constrain the parameter
$\Lambda$ for the same model as in previous section. Last section
contains our conclusion.

\section{$f(R,L_{m})$ gravity}
A maximal extension of the Hilbert-Einstein action for $f(R)$ gravity has been proposed
in Ref.\cite{13}. Its action takes the following form
\begin{equation}\label{1}
S=\int f(R,L_{m})\sqrt{-g}d^{4}x,
\end{equation}
where $f(R,L_{m})$ is an arbitrary function of the Ricci scalar $R$, and of the Lagrangian density
corresponding to matter, $L_{m}$. When $f(R,L_{m})=\dfrac{1}{2}f_{1}(R)+G(L_{m})f_{2}(R)$, the action of $f(R)$ gravity with arbitrary
matter¨Cgeometry coupling are recovered, where $f_{i}(R)$ ($i = 1,
2$) and $G(L_{m})$ are arbitrary functions of the Ricci scalar
and the Lagrangian density of matter, respectively. Furthermore, by setting $f_{2}(R)=1+\lambda f_{2}(R)$, $G(L_{m})=L_{m}$; $f_{1}(R)=f(R)$, $f_{2}(R)=1$,
$G(L_{m})=L_{m}$ and $f_{1}(R)=R$, $f_{2}(R)=1$ and
$G(L_{m})=L_{m}$, action (\ref{1}) can be reduced to the
context of $f(R)$ gravity with non-minimal coupling, non-coupling between matter and geometry and GR, respectively.

Varying the action (\ref{1}) with respect to the metric $g^{\mu\nu}$
yields the field equations
\begin{equation}\label{2}
\begin{array}{rcl}
&
&f_{R}(R,L_{m})R_{\mu\nu}+(g_{\mu\nu}\square-\triangledown_{\mu}\triangledown_{\nu})f_{R}(R,L_{m})-\dfrac{1}{2}[f(R,L_{m})-f_{L_{m}}(R,L_{m})L_{m}]g_{\mu\nu}\\&
&=\dfrac{1}{2}f_{L_{m}}(R,L_{m})T_{\mu\nu},
\end{array}
\end{equation}
where $\Box=g^{\mu\nu}\nabla_{\mu}\nabla_{\nu}$, $f_{R}(R,L_{m})=\partial f(R,L_{m})/\partial R$ and $f_{L_{m}}(R,L_{m})=\partial f(R,L_{m})/\partial L_{m}$, respectively.
To obtain Eq.(\ref{2}), it has been assumed that the Lagrangian density of matter, $L_{m}$, only depended on the components of the metric
tensor and not on its derivatives. The matter energy-momentum tensor is defined as
\begin{equation}\label{3}
T_{\mu\nu}=-\frac{2}{\sqrt{-g}}\frac{\delta(\sqrt{-g}L_{m})}{\delta
g^{\mu\nu}}.
\end{equation}

The contraction of Eq.(\ref{2}) provides
\begin{equation}\label{4}
\begin{array}{rcl}
&
&f_{R}(R,L_{m})R+3\Box f_{R}(R,L_{m})-2[f(R,L_{m})-f_{L_{m}}(R,L_{m})L_{m}]\\&
&=\dfrac{1}{2}f_{L_{m}}(R,L_{m})T,
\end{array}
\end{equation}
where $T=T^{\mu}_{\mu}$.

Taking the covariant divergence of Eq.(\ref{2}) and using the
mathematical identity\cite{22},
\begin{equation}\label{5}
\nabla^{\mu}[f_{R}(R,L_{m})R_{\mu\nu}-\dfrac{1}{2}f(R,L_{m})g_{\mu\nu}+(g_{\mu\nu}\square-\triangledown_{\mu}\triangledown_{\nu})f_{R}(R,L_{m})]\equiv0,
\end{equation}
one deduces the following generalized covariant conservation equation
\begin{equation}\label{6}
\nabla^{\mu}T_{\mu\nu}=2\nabla^{\mu}\ln[f_{L_{m}}(R,L_{m})]\frac{\partial L_{m}}{\partial g^{\mu\nu}}.
\end{equation}
It is clear that the non-minimal coupling between curvature and
matter yields a non-trivial exchange of energy and momentum between
the geometry and matter fields\cite{23}. However, once the $L_{m}$
is given, by choosing appropriate forms of $f_{L_{m}}(R,L_{m})$, one
can construct, at least in principle, conservative model in
$f(R,L_{m})$ gravity.

\section{Energy conditions in $f(R,L_{m})$ gravity}
\subsection{The Raychaudhuri equation}
In order to obtain the energy conditions in $f(R,L_{m})$ gravity, we
simply review the Raychaudhuri equation which is the physical origin
of the NEC and the SEC\cite{24}.

In the case of a congruence of timelike geodesics defined by the
vector field $u^{\mu}$, the Raychaudhuri equation is given by
\begin{equation}\label{10}
\frac{d\theta}{d\tau}=-\dfrac{1}{3}\theta^{2}-\sigma_{\mu\nu}\sigma^{\mu\nu}+\omega_{\mu\nu}\omega^{\mu\nu}-R_{\mu\nu}u^{\mu}u^{\nu},
\end{equation}
where $R_{\mu\nu}$, $\theta$, $\sigma_{\mu\nu}$ and
$\omega_{\mu\nu}$ are the Ricci tensor, the expansion parameter, the
shear and the rotation associated with the congruence, respectively.
While in the case of a congruence of null geodesics defined by the
vector field $k^{\mu}$, the Raychaudhuri equation is given by
\begin{equation}\label{11}
\frac{d\theta}{d\tau}=-\dfrac{1}{2}\theta^{2}-\sigma_{\mu\nu}\sigma^{\mu\nu}+\omega_{\mu\nu}\omega^{\mu\nu}-R_{\mu\nu}k^{\mu}k^{\nu}.
\end{equation}

From above expressions, it is clear that the Raychaudhuri equation
is purely geometric and independent of gravity theory. In order
to constrain the energy-momentum tensor by the Raychaudhuri
equation, one can use the Ricci tensor from the field equations of
gravity to make a connection. Namely, through the combination of the
field equations of gravity and the Raychaudhuri equation, one can
obtain physical conditions for the energy-momentum tensor. Since
$\sigma^{2}\equiv\sigma_{\mu\nu}\sigma^{\mu\nu}\geq0$ (the shear is
a spatial tensor) and $\omega_{\mu\nu}=0$ (hypersurface orthogonal
congruence), from Eqs. (\ref{10}) and (\ref{11}), the conditions for
gravity to remain attractive ($d\theta/d\tau<0$) are
\begin{equation}\label{12}
R_{\mu\nu}u^{\mu}u^{\nu}\geq0  ~~~~~~~~~~SEC,
\end{equation}
\begin{equation}\label{13}
R_{\mu\nu}k^{\mu}k^{\nu}\geq0  ~~~~~~~~~~NEC.
\end{equation}

Thus by means of the relationship (\ref{12}) and the Einstein's field equations,
one obtains
\begin{equation}\label{14}
R_{\mu\nu}u^{\mu}u^{\nu}=(T_{\mu\nu}-\dfrac{T}{2}g_{\mu\nu})u^{\mu}u^{\nu}\geq0,
\end{equation}
where $T_{\mu\nu}$ is the energy-momentum tensor and $T$ is its
trace. If one considers a perfect fluid with energy density $\rho$
and pressure $p$,
\begin{equation}\label{15}
T_{\mu\nu}=(\rho+p)U_{\mu}U_{\nu}-pg_{\mu\nu},
\end{equation}
the relationship (\ref{14}) turns into the well-known SEC of the
Einstein's theory, i.e.,
\begin{equation}\label{16}
\rho+3p\geq0.
\end{equation}

Similarly, by using the relationship (\ref{13}) and the Einstein's
field equations, one has
\begin{equation}\label{17}
T_{\mu\nu}k^{\mu}k^{\nu}\geq0.
\end{equation}
Then considering Eq.(\ref{15}), the familiar NEC of GR can be reproduced as:
\begin{equation}\label{18}
\rho+p\geq0.
\end{equation}

\subsection{Energy conditions in $f(R,L_{m})$ gravity}
The Einstein
tensor resulting from the field equations (\ref{2}) is
\begin{equation}\label{19}
G_{\mu\nu}\equiv R_{\mu\nu}-\dfrac{1}{2}g_{\mu\nu}R = T_{\mu\nu}^{eff},
\end{equation}
where the effective energy-momentum tensor $T_{\mu\nu}^{eff}$ is
defined as follows:
\begin{equation}\label{20}
\begin{array}{rcl}
T_{\mu\nu}^{eff}&=&\dfrac{1}{f_{R}(R,L_{m})}\bigg\{\dfrac{1}{2}g_{\mu\nu}[f(R,L_{m})-Rf_{R}(R,L_{m})]\\&
&-(g_{\mu\nu}\square-\triangledown_{\mu}\triangledown_{\nu})f_{R}(R,L_{m})+\dfrac{1}{2}f_{L_{m}}(R,L_{m})L_{m}g_{\mu\nu}\\ & &+\dfrac{1}{2}f_{L_{m}}(R,L_{m})T_{\mu\nu}\bigg\},
\end{array}
\end{equation}
where $\Box=g^{\mu\nu}\nabla_{\mu}\nabla_{\nu}$, $f_{R}(R,L_{m})=\partial f(R,L_{m})/\partial R$ and $f_{L_{m}}(R,L_{m})=\partial f(R,L_{m})/\partial L_{m}$, respectively.
Contracting Eq.(\ref{20}), we have
\begin{equation}\label{21}
\begin{array}{rcl}
T^{eff} & = &\dfrac{1}{f_{R}(R,L_{m})}\bigg\{2[f(R,L_{m})-Rf_{R}(R,L_{m})]-3\Box f_{R}(R,L_{m})\\ & &+2f_{L_{m}}(R,L_{m})L_{m}+\dfrac{1}{2}f_{L_{m}}(R,L_{m})T\bigg\},
\end{array}
\end{equation}
where $T=g^{\mu\nu}T_{\mu\nu}$. Thus, we can write the Ricci tensor in
terms of an effective stress-energy tensor and its trace, i.e.,
\begin{equation}\label{22}
R_{\mu\nu}=T_{\mu\nu}^{eff}-\dfrac{1}{2}g_{\mu\nu}T^{eff}.
\end{equation}

In order to keep gravity attractive, besides the expressions
(\ref{12}) and (\ref{13}), the following additional condition should
be required
\begin{equation}\label{23}
\dfrac{f_{L_{m}}(R,L_{m})}{f_{R}(R,L_{m})} > 0.
\end{equation}
Note that this condition is independent of the ones derived from the
Raychaudhuri equation (i.e., the expressions (\ref{12}) and
(\ref{13})), and only relates to an effective gravitational coupling.

The FRW metric is chosen as:
\begin{equation}\label{24}
ds^{2}=dt^{2}-a^{2}(t)ds^{2}_{3},
\end{equation}
where $a(t)$ is the scale factor and $ds^{2}_{3}$ contains the
spacial part of the metric. Using this metric, we can obtain
$R=-6(2H^{2}+\dot{H})$, where $H=\dot{a}(t)/a(t)$ is the Hubble
expansion parameter, and
$\Gamma^{0}_{\mu\nu}=a(t)\dot{a}(t)\delta_{\mu\nu} ~(\mu, \nu\neq0)$,
which are the components of the affine connection.

By using the relationship (\ref{12}) and Eq. (\ref{22}), the SEC can
be given as:
\begin{equation}\label{25}
T_{\mu\nu}^{eff}u^{\mu}u^{\nu}-\dfrac{1}{2}T^{eff}\geq0,
\end{equation}
where we have used the condition $g_{\mu\nu}u^{\mu}u^{\nu}=1$.
Taking the energy-momentum tensor, $T_{\mu\nu}$, to be a perfect fluid
(i.e., Eq.(\ref{15})) and considering the condition (\ref{23}), Eq.(\ref{25}) turns into
\begin{equation}\label{26}
\begin{array}{rcl}& &
\rho+3p-\dfrac{2}{f_{L_{m}}(R,L_{m})}\Big[f(R,L_{m})-Rf_{R}(R,L_{m})\Big]\\& &+\dfrac{6}{f_{L_{m}}(R,L_{m})}\Big[f_{RRR}(R,L_{m})\dot{R}^{2}
+f_{RR}(R,L_{m})\ddot{R}\\& &+Hf_{RR}(R,L_{m})\dot{R}\Big]-2L_{m}\geq0,
\end{array}
\end{equation}
where the dot denotes differentiation with respect to cosmic time.
This is the SEC in $f(R,L_{m})$ gravity.

The NEC in $f(R,L_{m})$ gravity can be expressed as:
\begin{equation}\label{27}
T_{\mu\nu}^{eff}k^{\mu}k^{\nu}\geq0.
\end{equation}
By the same method as the SEC, the above relationship can be changed
into
\begin{equation}\label{28}
\begin{array}{rcl}& &
\rho+p+\dfrac{2}{f_{L_{m}}(R,L_{m})}\Big[f_{RRR}(R,L_{m})(\dot{R})^{2}+f_{RR}(R,L_{m})\ddot{R}\Big]\geq0.
\end{array}
\end{equation}

From above discussion, it is worth stressing that by taking
$f(R,L_{m})=\dfrac{1}{2}f_{1}(R)+G(L_{m})f_{2}(R)$ in expressions
(\ref{26}) and (\ref{28}), we can obtain the SEC and the NEC in
$f(R)$ gravity with arbitrary matter¨Cgeometry coupling, which are
just the results given in Ref.\cite{19}. Furthermore, by setting
$f_{2}(R)=1+\lambda f_{2}(R)$, $G(L_{m})=L_{m}$ and $f_{1}(R)=f(R)$,
$f_{2}(R)=1$ and $G(L_{m})=L_{m}$, we can get the SEC and the NEC in
$f(R)$ gravity with non-minimal coupling and non-coupling between
matter and geometry, which are just the same as the ones in
Ref.\cite{18} and Ref.\cite{17}, respectively. For $f_{1}(R)=R$,
$f_{2}(R)=1$ and $G(L_{m})=L_{m}$, the SEC and the NEC in GR, i.e.,
$\rho+3p\geq0$ and $\rho+p\geq0$, can be reproduced.

Note that both above expressions of the SEC and the NEC are directly
derived from the Raychaudhuri equation. However, from the calculation, we find that the equivalent results can
be obtained by taking the transformations $ \rho \rightarrow
\rho^{eff}$ and $p\rightarrow p^{eff}$ into $\rho+3p\geq0$ and
$\rho+p\geq0$, respectively. Thus by extending this approach to $\rho-p\geq0$ and
$\rho\geq0$, we can give the DEC and the WEC in $f(R,L_{m})$ gravity.

By means of Eqs. (\ref{20}) and (\ref{24}), the effective energy
density and the effective pressure can be derived as:
\begin{equation}\label{29}
\begin{array}{rcl}
\rho^{eff} & = &\dfrac{1}{f_{R}(R,L_{m})}\Big\{\dfrac{1}{2}[f(R,L_{m})-Rf_{R}(R,L_{m})]-3Hf_{RR}(R,L_{m})\dot{R}\\& &
+\dfrac{1}{2}f_{L_{m}}(R,L_{m})L_{m}+\dfrac{1}{2}f_{L_{m}}(R,L_{m})\rho\Big\},
\end{array}
\end{equation}

\begin{equation}\label{30}
\begin{array}{rcl}
p^{eff} & = &\dfrac{1}{f_{R}(R,L_{m})}\Big\{\dfrac{1}{2}[Rf_{R}(R,L_{m})-f(R,L_{m})]+f_{RRR}(R,L_{m})\dot{R}^{2}\\& &
+f_{RR}(R,L_{m})\ddot{R}+3Hf_{RR}(R,L_{m})\dot{R}-\dfrac{1}{2}f_{L_{m}}(R,L_{m})L_{m}\\& &+\dfrac{1}{2}f_{L_{m}}(R,L_{m})p\Big\}.
\end{array}
\end{equation}
Then the corresponding DEC and WEC in $f(R,L_{m})$ gravity can be respectively given as:
\begin{equation}\label{31}
\begin{array}{rcl}& &
\rho-p+\dfrac{2}{f_{L_{m}}(R,L_{m})}\Big[f(R,L_{m})-Rf_{R}(R,L_{m})\Big]\\ & &-\dfrac{2}{f_{L_{m}}(R,L_{m})}\Big[f_{RRR}(R,L_{m})\dot{R}^{2}
+f_{RR}(R,L_{m})\ddot{R}\\& &+6Hf_{RR}(R,L_{m})\dot{R}\Big]+2L_{m}\geq0,
\end{array}
\end{equation}

\begin{equation}\label{32}
\begin{array}{rcl}& &
\rho+\dfrac{1}{f_{L_{m}}(R,L_{m})}\Big[f(R,L_{m})-Rf_{R}(R,L_{m})\Big]\\& &-\dfrac{6}{f_{L_{m}}(R,L_{m})}Hf_{RR}(R,L_{m})\dot{R}
+L_{m}\geq0.
\end{array}
\end{equation}

We emphasize that by taking
$f(R,L_{m})=\dfrac{1}{2}f_{1}(R)+G(L_{m})f_{2}(R)$, above
expressions are the DEC and the WEC in $f(R)$ gravity with arbitrary
matter¨Cgeometry coupling, which are just the same as the ones in
Ref.\cite{19}. Furthermore, by setting $f_{2}(R)=1+\lambda
f_{2}(R)$, $G(L_{m})=L_{m}$ and $f_{1}(R)=f(R)$, $f_{2}(R)=1$ and
$G(L_{m})=L_{m}$, the results given by us are the DEC and the WEC in
$f(R)$ gravity with non-minimal coupling and non-coupling between
matter and geometry, which are consistent with the results given in
Ref.\cite{18} and Ref.\cite{17}, respectively. For $f_{1}(R)=R$,
$f_{2}(R)=1$ and $G(L_{m})=L_{m}$, the DEC  and the WEC in GR, i.e.,
$\rho-p\geq0$ and $\rho\geq0$, can be reproduced.

In order to get some insights on the meaning of these energy
conditions, we consider a specific type of model in $f(R, L_{m})$ gravity, which takes the form as
\begin{equation}\label{33}
f(R, L_{m})=\Lambda\exp(\dfrac{1}{2\Lambda}R+\dfrac{1}{\Lambda}L_{m}),
\end{equation}
where $\Lambda>0$ is an arbitrary constant. In the limit $(1/2\Lambda)R+(1/\Lambda)L_{m}\ll1$, we obtain
\begin{equation}\label{Limit}
f(R, L_{m})\approx \Lambda+\dfrac{R}{2}+L_{m}+...,
\end{equation}
which is the Lagrangian density of GR with a cosmological constant.

In the FRW cosmology, the energy conditions for model of Eq.(\ref{33}) can be written as
\begin{equation}\label{34}
A+B\geq C
\end{equation}
where $A$, $B$ and $C$ depend on the energy condition under
study. For the SEC, one finds
\begin{subequations}
\begin{equation}\label{SEC,A}
A^{SEC}=\rho+3p,
\end{equation}

\begin{equation}\label{SEC,B}
B^{SEC}=R+\dfrac{3H\dot{R}}{2\Lambda}+\dfrac{3(\dot{R})^{2}\ddot{R}\exp(\dfrac{R+2L_{m}}{2\Lambda})}{16\Lambda^{3}},
\end{equation}

\begin{equation}\label{SEC,C}
C^{SEC}=2(\Lambda+L_{m}).
\end{equation}
\end{subequations}

For the NEC, one obtains
\begin{subequations}
\begin{equation}\label{NEC,A}
A^{NEC}=\rho+p,
\end{equation}

\begin{equation}\label{NEC,B}
B^{NEC}=\dfrac{\dot{R}^{2}\ddot{R}\exp(\dfrac{R+2L_{m}}{2\Lambda})}{16\Lambda^{3}},
\end{equation}

\begin{equation}\label{NEC,C}
C^{NEC}=0.
\end{equation}
\end{subequations}

For the DEC, one has
\begin{subequations}
\begin{equation}\label{DEC,A}
A^{DEC}=\rho-p,
\end{equation}

\begin{equation}\label{DEC,B}
B^{DEC}=-\left[R+\dfrac{3H\dot{R}}{\Lambda}+\dfrac{\dot{R}^{2}\ddot{R}\exp(\dfrac{R+2L_{m}}{2\Lambda})}{16\Lambda^{3}}\right],
\end{equation}

\begin{equation}\label{DEC,C}
C^{DEC}=-2(\Lambda+L_{m}).
\end{equation}
\end{subequations}

Finally, for the WEC, one gets
\begin{subequations}
\begin{equation}\label{WEC,A}
A^{WEC}=\rho,
\end{equation}

\begin{equation}\label{WEC,B}
B^{WEC}=-\dfrac{1}{2}(R+\dfrac{3H\dot{R}}{\Lambda}),
\end{equation}

\begin{equation}\label{WEC,C}
C^{WEC}=-(\Lambda+L_{m}).
\end{equation}
\end{subequations}
Given these definitions, the study of all the energy conditions can
be performed by satisfying the inequality (\ref{34}).

\section{Constraints on $f(R, L_{m})$ gravity}
The Ricci scalar $R$ and its derivatives can be expressed by the
parameters of the deceleration $(q)$, the jerk $(j)$ and the snap
$(s)$\cite{25}, namely,
\begin{subequations}
\begin{equation}\label{qjs,a}
R=-6H^{2}(1-q),
\end{equation}
\begin{equation}\label{qjs,b}
\dot{R}=-6H^{3}(j-q-2),
\end{equation}
\begin{equation}\label{qjs,c}
\ddot{R}=-6H^{4}(s+q^{2}+8q+6),
\end{equation}
\end{subequations}
where
\begin{equation}\label{qjs}
q=-\dfrac{1}{H^{2}}\dfrac{\ddot{a}}{a},
~~j=\dfrac{1}{H^{3}}\dfrac{\stackrel{...}{a}}{a}, ~~and
~~s=\dfrac{1}{H^{4}}\dfrac{\stackrel{....}{a}}{a}.
\end{equation}
Thus, the energy conditions (\ref{26}), (\ref{28}), (\ref{31}) and
(\ref{32}) can be rewritten as:

\begin{subequations}
\begin{equation}\label{qjs,SEC}
\begin{array}{rcl}
& &\rho+3p-2\Big[\dfrac{f(R,L_{m})-6H^{2}(q-1)f_{R}(R,L_{m})}{f_{L_{m}}(R,L_{m})}+L_{m}\Big]\\& &
+[-1-36H^{6}(j-q-2)(6+8q+q^{2}+s)f_{RRR}(R,L_{m})]\times\\& &
\dfrac{36H^{4}(j-q-2)f_{RR}(R,L_{m})}{f_{L_{m}}(R,L_{m})}
\geq0, (SEC)
\end{array}
\end{equation}

\begin{equation}\label{qjsNEC}
\begin{array}{rcl}
& &\rho+p-432H^{10}(2-j+q)^{2}(6+8q+q^{2}+s)
\times\\& &\dfrac{f_{RR}(R,L_{m})f_{RRR}(R,L_{m})}{f_{L_{m}}(R,L_{m})}\geq0, (NEC)
\end{array}
\end{equation}

\begin{equation}\label{qjs,DEC}
\begin{array}{rcl}
& &\rho-p+2\dfrac{f(R,L_{m})-6H^{2}(q-1)f_{R}(R,L_{m})}{f_{L_{m}}(R,L_{m})}+2L_{m}\\& &
-[-1-6H^{6}(j-q-2)(6+8q+q^{2}+s)f_{RRR}(R,L_{m})]\times\\& &\dfrac{72H^{4}(j-q-2)f_{RR}(R,L_{m})}{f_{L_{m}}(R,L_{m})}\geq0, (DEC)
\end{array}
\end{equation}

\begin{equation}\label{qjs,WEC}
\begin{array}{rcl}
& &\rho+\dfrac{f(R,L_{m})-6H^{2}(q-1)f_{R}(R,L_{m})}{f_{L_{m}}(R,L_{m})}\\& &
+36H^{4}(j-q-2)\dfrac{f_{RR}(R,L_{m})}{f_{L_{m}}(R,L_{m})}+L_{m}\geq0. (WEC)
\end{array}
\end{equation}
\end{subequations}

To exemplify how to use these energy conditions to constrain
$f(R,L_{m})$ gravity, we consider the same model as given in Eq.(\ref{33}). Since there has been no reliable
measurement for the parameter of the snap  $(s)$ up to now, we only focus on
the WEC. Under the requirement $f'(R)>0 $ for all $R$, the WEC (\ref{qjs,WEC}) in this particular case
is
\begin{equation}\label{WEC,jq}
\rho + \Lambda - 3H^{2}(q-1) + \dfrac{9H^{4}(j - q - 2)}{\Lambda} +
L_{m} \geq 0.
\end{equation}

Furthermore, by taking $H_{0} = 73.8$\cite{26}, $q_{0} = -0.81 \pm
0.14$ and $j_{0} = 2.16^{+ 0.81}_{- 0.75}$\cite{27} (the subscript
$0$ denotes the present value) and considering two specific forms of
Lagrangian density of matter, i.e., $L_{m} = -\rho$, where $\rho$ is
the energy density and $L_{m} = p$, where $p$ is the
pressure\cite{21}, respectively, the results of expression
(\ref{WEC,jq}) are as follows: for $L_{m} = -\rho$, the parameter
$\Lambda$ can be taken as any real number. Considering $\Lambda > 0$
given in Eq.(\ref{33}), the range of the parameter $\Lambda$ is
$\Lambda > 0$. While for $L_{m} = P$, the parameter $\Lambda$ can be
taken as any positive real number, where we have used $p = 0$ for
pressureless matter and $\Omega_{m0} = 0.27$ in calculation.

Note that although the range of the parameter $\Lambda$ obtained
from above discussions is the same in both particular cases, there
is no equivalence between the two specific forms of Lagrangian
density of matter. This is because the two specific forms of
Lagrangian density of matter have different forms of the
gravitational field equations.

\section{conclusion}
In this paper, we have derived the energy conditions (SEC, NEC, DEC, WEC)
in $f(R,L_{m})$ gravity. For the SEC and the NEC, the Raychaudhuri equation which
is the physical origin of them has been used. From the calculation,
we found that the equivalent results can be obtained by taking the
transformations $ \rho \rightarrow \rho^{eff}$ and $p\rightarrow
p^{eff}$ into $\rho+3p\geq0$ and $\rho+p\geq0$, respectively. Thus by extending
this approach to $\rho-p\geq0$ and $\rho\geq0$, the DEC and the WEC
in $f(R,L_{m})$ gravity were obtained. The condition of keeping gravity
attractive was also found but the approach of getting it was
different.

It is worth stressing that the energy conditions obtained in this
paper are quite general, which includes the corresponding results
given in $f(R)$ theories of gravity with arbitrary coupling, non-minimal coupling and non-coupling between matter and geometry as well as in GR, respectively,
as special cases.

Furthermore, to exemplify how to use these derived energy conditions
to constrain $f(R,L_{m})$ gravity, we considered a special model
with $f(R,
L_{m})=\Lambda\exp(\dfrac{1}{2\Lambda}R+\dfrac{1}{\Lambda}L_{m})$.
By virtue of the WEC, present astronomical observations and
considering $L_{m}=-\rho$ and $L_{m}=p$, respectively, we
constrained the range of the parameter $\Lambda$.

\section*{Acknowledgements}
The authors would like to thank professor Zong-Hong Zhu
for discussions. This work was supported by the National Natural Science Foundation of
China under the Distinguished Young Scholar Grant 10825313,
the Ministry of Science and Technology national basic science Program (Project 973)
under Grant No.2012CB821804, the Fundamental Research Funds for the Central
Universities and Scientific Research Foundation of Beijing Normal University.

%f_{R}(R,L_{m})R_{\mu\nu}+(g_{\mu\nu}\square-\triangledown_{\mu}\triangledown_{\nu})f_{R}(R,L_{m})-\dfrac{1}{2}[f(R,L_{m})-f_{L_{m}}(R,L_{m})L_{m}]g_{\mu\nu}\\&
%&=\dfrac{1}{2}f_{L_{m}}(R,L_{m})T_{\mu\nu},

%\rho+3p-\dfrac{2}{f_{L_{m}}(R,L_{m})}[f(R,L_{m})-Rf_{R}(R,L_{m})]+\dfrac{6}{f_{L_{m}}(R,L_{m})}[f_{RRR}(R,L_{m})(\dot{R})^{2}\\& &
%+f_{RR}(R,L_{m})\ddot{R}+Hf_{RR}(R,L_{m})\dot{R}]-2L_{m}\geq0,

\end{document}